%% file: paper.tex
\pdfoutput=1

\documentclass[conference]{IEEEtran}
\usepackage[pdftex]{graphicx}

\ifCLASSINFOpdf
\else
\fi
\begin{document}
%
\title{Fast Processing of Large Graph Applications Using Asynchronous Architecture\vspace{-.2in}}

\author{\IEEEauthorblockN{Michel A. Kinsy, Rashmi S. Agrawal and Hien D. Nguyen}
\IEEEauthorblockA{
Adaptive and Secure Computing Systems (ASCS) Laboratory\\
Department of Electrical and Computer Engineering, Boston University\\
8 Saint Mary's Street, Boston MA 02215 - Email: mkinsy@bu.edu}
\vspace{-.4in} }


%


\maketitle

\begin{abstract}
Graph algorithms and techniques are increasingly being used in scientific and commercial applications to express relations and explore large data sets. Although conventional or commodity computer architectures, like CPU or GPU, can compute fairly well dense graph algorithms, they are often inadequate in processing large sparse graph applications. Memory access patterns, memory bandwidth requirements and on-chip network communications in these applications do not fit in the conventional program execution flow. 
In this work, we propose and design a new architecture for fast processing of large graph applications. To leverage the lack of the spatial and temporal localities in these applications and to support scalable computational models, we design the architecture around two key concepts. (1) The architecture is a multicore processor of independently clocked processing elements. These elements communicate in a self-timed manner and use handshaking to perform synchronization, communication, and sequencing of operations. By being asynchronous, the operating speed at each processing element is determined by actual local latencies rather than global worst-case latencies. We create a specialized ISA to support these operations. (2) The application compilation and mapping process uses a graph clustering algorithm to optimize parallel computing of graph operations and load balancing. Through the clustering process, we make scalability an inherent property of the architecture where task-to-element mapping can be done at the graph node level or at node cluster level. A prototyped version of the architecture outperforms a comparable CPU by 10$\sim$20x across all benchmarks and provides 2$\sim$5x better power efficiency when compared to a GPU.
\end{abstract}


%
\IEEEpeerreviewmaketitle

\input{intro}
\input{arch}

\input{eval}

\bibliographystyle{IEEEtran}

\end{document}

%% file: intro.tex
\section{Introduction}
\label{sec:intro}
Advances in mobile computing, coupled with the proliferation of online social networks, have given rise to a new class of applications and computing challenges~\cite{Kyrola}. These applications tend to be relational by nature. In other words, they express or encode relations, communications, connectivity and interactions between people, places, objects or systems. As such, the data of interest in these applications are often best represented in the form of graphs. Graph-based applications range from social network analyses to anomaly detections~\cite{lumsdaine}. 
For computing purposes, graphs are commonly represented in one of two forms: (1) as adjacency matrix or (2) as adjacency list. Adjacency Matrix works well for densely connected graphs, i.e., the number of edges in the graph is close to the maximal number of edges. In general, computing on dense graphs can be easily parallelized and GPU and SIMD architectures have proven to be the platform of choice for executing such graph-based applications~\cite{lumsdaine}. Unfortunately, the vast majority of large graph-based applications are sparse. For the efficient storage of large sparse graphs, adjacency list or other compressed representation schemes are used.  Memory access and load balancing are some of the key bottlenecks to the efficient processing of large sparse graph algorithms and applications~\cite{Kyrola}. The memory access patterns often lack spatial and temporal localities resulting in high cache miss rates. Current cache-based processor architectures are simply not well suited for the computational flow of graph processing. 
In addition to the storage problem, computing on large sparse graphs currently presents a number of challenges including effective programming abstractions and models of computation that leverage the graph structure in the application. 
%
In this work, we present a domain-specific architecture tailored to graph-based algorithms and applications. 

%% file: arch.tex
\section{Proposed Graph Processor Architecture}
\label{sec:arch}

\begin{figure}[!ht]
\vspace{-.1in}
\begin{center}
\includegraphics[width=3.25in]{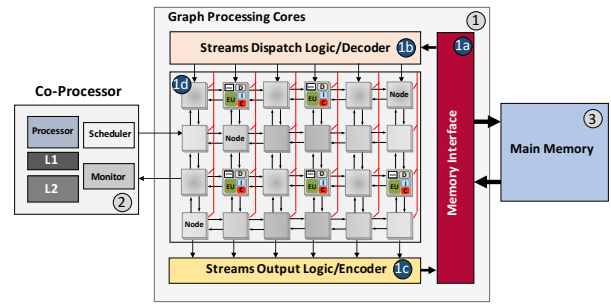}
\vspace{-.15in}
\caption{Proposed graph processor system architecture.}
\label{fig:arch}
\end{center}
\vspace{-.2in}
\end{figure}

Figure~\ref{fig:arch} shows an illustration of the proposed architecture. The three key modules of the architecture are (1) the graph processor, (2) the co-processor and (3) the main memory. The graph processor (1) module has a Memory Interface unit (1a) to coordinate batch accesses to the main memory or external memory units, a Dispatch Logic (1b) to perform scatter operations on data from the main memory, an Output Logic (1c) to gather output data from the graph processor, and a systolic array of simple processing elements called \textit{Node Arithmetic Logic Engines} (NALEs) (1d) to carry out the actual graph computations. The co-processor (2) performs three key functions. It (1) executes non-graph parts of the application, (2) schedules the graph part of the application and (3) monitors the execution flow of the graph.

\begin{figure}[!ht]
\vspace{-.2in}
\begin{center}
\includegraphics[width=2.0in]{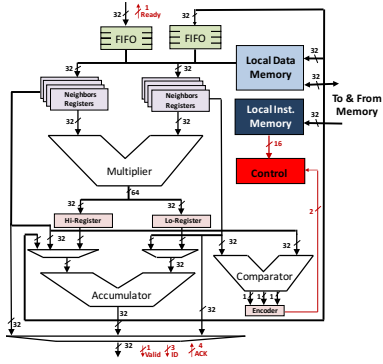}
\vspace{-.1in}
\caption{Micro-architecture of a Node Arithmetic Logic Engine (NALE).}
\label{fig:micro}
\end{center}
\vspace{-.3in}
\end{figure}

\textit{Graph Processor Micro-Architecture}: 
Figure~\ref{fig:micro} shows the micro-architecture of a NALE.  The NALE is optimized for fast MAC (Multiply-And-Accumulate) operations with a three-state output comparator for fast node value sorting. It has two FIFO structures, one to communicate with neighbors and one internal FIFO to emulate multiple graph nodes (node cluster mode execution). 


\begin{figure}[!ht]
\vspace{-.1in}
\begin{center}
\includegraphics[width=3.5in]{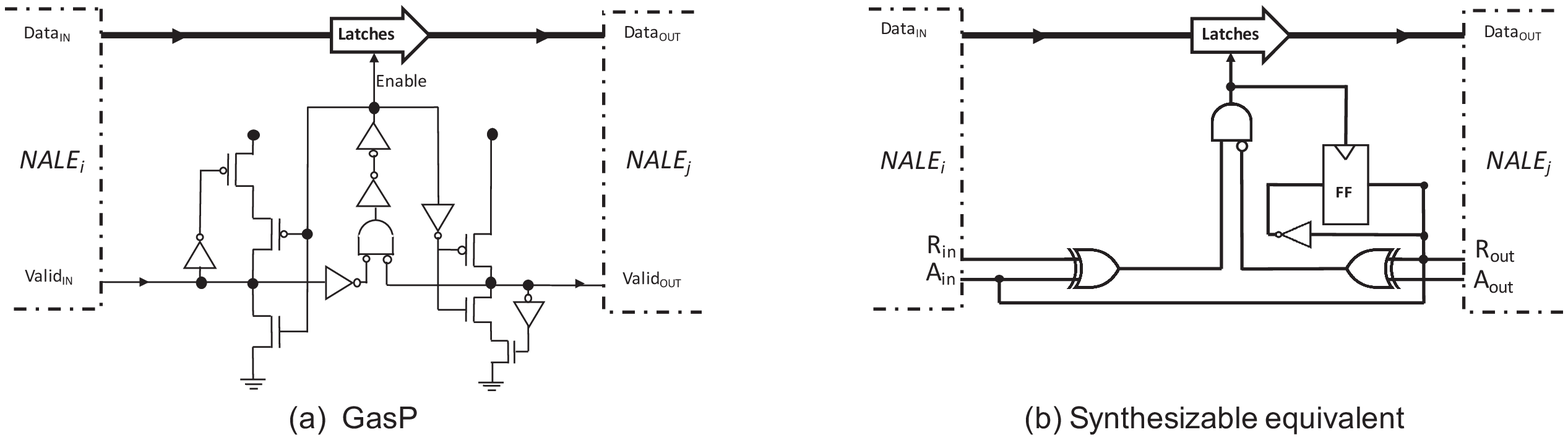}
\vspace{-.25in}
\caption{GasP asynchronous communication circuit between NALEs.}
\label{fig:gasp}
\end{center}
\vspace{-.2in}
\end{figure}

Each NALE operates independently of others depending on the readiness of inputs. Communicating through FIFOs only allows each NALE to run on its own clock speed. Furthermore, this approach allows us to adopt a GasP asynchronous \cite{Roncken} design methodology that can seamlessly scale to hundreds of thousands of NALEs. Figure~\ref{fig:gasp}(a) illustrates the the clockless handshake logic between NALEs. In addition to the scalability benefits, the absence of a global clock allows for the underlining data dependencies to dictate application execution time. Figure~\ref{fig:gasp}(b) shows the synthesizable equivalent of the GasP circuit. 
%

\begin{figure}[!ht]
\vspace{-.2in}
\begin{center}
\includegraphics[width=3.0in]{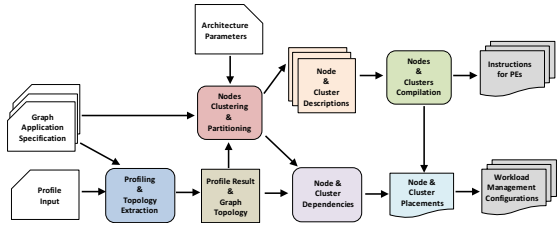}
\vspace{-.1in}
\caption{Compilation process steps for the graph processor.}
\label{fig:compile}
\end{center}
\vspace{-.3in}
\end{figure}

\textit{Model of computation and compilation }: 
An asynchronous model of computation is adopted to fully take advantage of the graph processor. Given a graph application specification and a number of available NALEs for its computation, the execution preprocessing flow follows five key steps. Figure~\ref{fig:compile} illustrates these steps.  In the first step, the application is profiled to extract the graph topology, followed by the clustering of nodes, clusters dependency analysis, placement and finally the compilation step. 

%% file: eval.tex
\section{Architecture Evaluation}
\label{sec:eval}
\textit{Experimental setup}: 
To get high-fidelity performance and power measurements for the proposed architecture, we prototype it alongside a conventional CPU and a GPU with comparable complexity in FPGA. The Xilinx Virtex7-XC7VX980T FPGA device is used for our prototyping platform. We implement a synthesizable RTL version of the graph processor. We use the 7-stage RISC core in the Heracles~\cite{heracles} RTL simulator for the CPU. We adopte the MIAOW open-source general-purpose graphics processor (GPGPU) based on the AMD Southern Islands ISA~\cite{Balasubramanian} for the GPU architecture. The three architectures are evaluated based on their execution time and power. 

\textit{Graph algorithms and applications}: 
For the evaluation, we consider a set of representative graph algorithms, namely, Single Source Shortest Path (SSSP), Breadth First Search (BFS), Depth First Search (DFS), PageRank (PR), Minimal Enclosing Triangles (MiniTri), and Connected Components (CC). 
%
We use three difference graph applications: (1) California road network (CA) which has 1,965,206 vertices, 2,766,607 edges and an average degree of 1.41, (2) Facebook social network (FB) with 2,937,612 vertices, 41,919,708 edges and an average degree of 14.3 and (3) Livejournal social network (LJ) with 4,847,571 vertices, 85,702,475 edges and 17.6 average degree. 

\begin{figure}[!ht]
\vspace{-.15in}
\begin{center}
\includegraphics[width=2.0in]{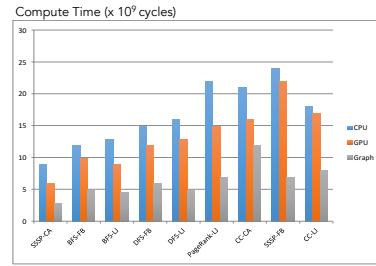}
\vspace{-.1in}
\caption{Performance in terms of execution for the three architecture types on the different graph applications.}
\vspace{-.1in}
\label{fig:perf1}
\end{center}
\vspace{-.125in}
\end{figure}

\begin{figure}[!ht]
\vspace{-.1in}
\begin{center}
\includegraphics[width=2.0in]{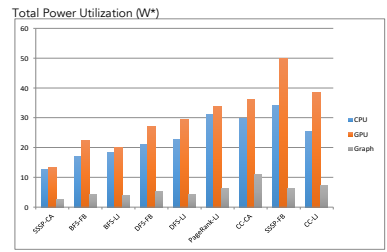}
\vspace{-.1in}
\caption{Efficiency in terms of power usage for the three architecture types on the different graph applications.}
\label{fig:perf2}
\end{center}
\vspace{-.25in}
\end{figure}

\textit{Results}: 
Figures~\ref{fig:perf1} and~\ref{fig:perf2} present the execution time in terms of number of cycles and power usage for each platform for the different applications and graph algorithms. 

%

